\documentclass{aastex631}

\hypersetup{linkcolor=red,citecolor=blue,filecolor=cyan,urlcolor=magenta}

\def\lessim{\mathrel{\hbox{\rlap{\hbox{\lower4pt\hbox{$\sim$}}}\hbox{$<$}}}}
\def\grtsim{\mathrel{\hbox{\rlap{\hbox{\lower4pt\hbox{$\sim$}}}\hbox{$>$}}}}

\usepackage{CJK}

\shorttitle{M31N 1926-07c}
\shortauthors{Shafter et al.}

\graphicspath{{./}{figures/}}

\begin{document}
\begin{CJK*}{UTF8}{gbsn}
\title{M31N 1926-07c: A Recurrent Nova in M31 with a 2.8 Year Recurrence Time}

\correspondingauthor{A. W. Shafter}
\email{ashafter@sdsu.edu}

\author[0000-0002-1276-1486]{Allen W. Shafter}
\affiliation{Department of Astronomy, San Diego State University, San Diego, CA 92182, USA}

\author[0000-0002-0835-225X]{Kamil Hornoch}
\affiliation{Astronomical Institute of the Czech Academy of Sciences, Fri\v{c}ova 298, CZ-251 65 Ond\v{r}ejov, Czech Republic}

\author[0000-0002-1330-1318]{Hana Ku\v{c}\'akov\'a}
\affiliation{Astronomical Institute of the Czech Academy of Sciences, Fri\v{c}ova 298, CZ-251 65 Ond\v{r}ejov, Czech Republic}
\affiliation{Research Centre for Theoretical Physics and Astrophysics, Institute of Physics, Silesian University in Opava, Bezru\v{c}ovo n\'am. 13, CZ-74601 Opava, Czech Republic}

\author[0000-0002-2770-3481]{Jingyuan Zhao (赵经远)}
\affiliation{Xingming Observatory, Mount Nanshan, Xinjiang, China}

\author[0000-0003-3005-2189]{Mi Zhang (张宓)}
\affiliation{Xingming Observatory, Mount Nanshan, Xinjiang, China}

\author[0000-0002-7292-3109]{Xing Gao (高兴)}
\affiliation{Xingming Observatory, Mount Nanshan, Xinjiang, China}

\author[0000-0003-0928-2000]{John Della Costa}
\affiliation{Department of Astronomy, San Diego State University, San Diego, CA 92182, USA}

\author[0000-0002-6023-7291]{William A. Burris}
\affiliation{Department of Astronomy, San Diego State University, San Diego, CA 92182, USA}

\author[0000-0003-4735-9128]{J. Grace Clark}
\affiliation{Department of Astronomy, San Diego State University, San Diego, CA 92182, USA}

\author[0000-0002-4387-6358]{Marek Wolf}
\affiliation{Astronomical Institute, Charles University, Prague, V Hole\v{s}ovi\v{c}k\'ach 2, CZ-180 00 Praha 8, Czech Republic}

\author[0000-0001-9383-7704]{Petr Zasche}
\affiliation{Astronomical Institute, Charles University, Prague, V Hole\v{s}ovi\v{c}k\'ach 2, CZ-180 00 Praha 8, Czech Republic}

\begin{abstract}

The M31 recurrent nova M31N 1926-07c has had five recorded eruptions. Well-sampled light curves of the two most recent outbursts, in January of 2020 (M31N 2020-01b) and September 2022 (M31N 2022-09a), are presented showing that the photometric evolution of the two events were quite similar, with peak magnitudes of $R=17.2\pm0.1$ and $R=17.1\pm0.1$, and $t_2$ times of $9.7\pm0.9$ and $8.1\pm0.5$ days for the 2020 and 2022 eruptions, respectively.
After considering the dates of the four most recent eruptions (where the cycle count is believed to be known), a mean recurrence interval of
$\langle P_\mathrm{rec}\rangle=2.78\pm0.03$ years is found, establishing that M31N 1926-07c has one of the shortest recurrence times known.

\end{abstract}


\section{Introduction} \label{sec:intro}

The nova M31N 1926-07c was first recognized to be recurrent by \citet{2015ApJS..216...34S} who discovered that two subsequent eruptions of the nova had been observed: one in October 1997 (M31N 1997-10f) and the other in August 2008 (M31N 2008-08b). Since the publication of that study just seven years ago, two additional eruptions have now been reported, the first in 2020 January (M31N 2020-01b), and the second in 2022 September (M31N 2022-09a).
As pointed out by \citet{2022ATel15609....1H,2022ATel15608....1H}, the latest two eruptions are significant in that they have revealed the recurrence time of this nova
to be much shorter than previously realized.

\section{The Mean Recurrence Time}

It is difficult to unambiguously establish the mean recurrence time for M31N 1926-07c
due to the large gap between the first recorded eruption in July of 1926 and October of 1997. However, if we restrict our analysis 
to the times of peak brightness for the four most recent observed eruptions, which have taken place over just the past $\sim$25 years, a conservative estimate for the mean recurrence time
is found to be $\langle P_\mathrm{rec}\rangle=2.78\pm0.03$ years. If we include the initial eruption from July of 1926 in the analysis,
we find that either $\langle P_\mathrm{rec}\rangle=2.75\pm0.01$ or $\langle P_\mathrm{rec}\rangle=2.83\pm0.01$ years, depending on whether 25 or 24 eruptions
were missed between July 1926 and October 1997.
The shorter mean recurrence time, $\langle P_\mathrm{rec}\rangle=2.75\pm0.01$ years, gives a marginally better fit to the data,
and is slightly favored.

Of all known recurrent novae, only M31N 2008-12a, with its
$0.99\pm0.02$ year mean inter-outburst interval \citep{2020AdSpR..66.1147D}, and possibly
M31N 2017-01e (= 2019-09d, 2022-03d) --
a recently-identified M31 recurrent nova with a likely mean eruption interval of $2.55\pm0.13$ years (Shafter et al. 2022, in preparation) --
have been observed to erupt more frequently.

\begin{figure*}
\includegraphics[angle=0,scale=0.7]{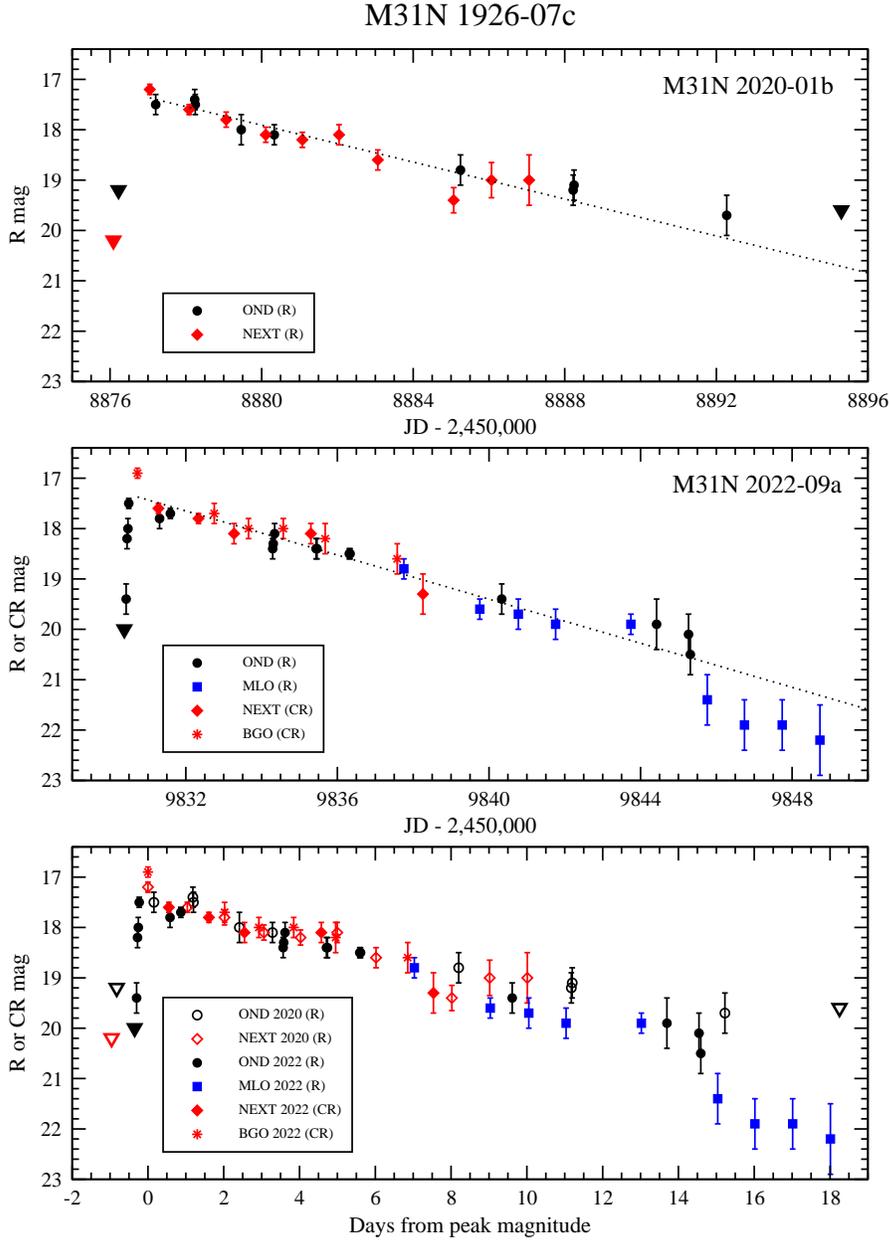}
\caption{The light curves for the 2020 January (M31N 2020-01b, top panel) and 2022 September (M31N 2022-09a, middle panel) eruptions of the recurrent nova M31N 1926-07c. The bottom panel shows both eruptions superimposed after adopting times of maximum of JD 2,458,877.048 and 2,459,830.725 for the 2020 and 2022 eruptions, respectively. Different symbols represent data taken with different telescopes as shown in the key (the 2020 data are shown as open symbols in the bottom panel). The downward-facing triangles represent lower limits on the magnitudes: NEXT (red) and OND (black). The BGO (NEXT) data for the 2022 eruption were taken through a luminance filter (no filter), and were calibrated using $R$-band secondary standard stars. The resulting data are given as $CR$ magnitudes. Only measurements taken within 14 days of maximum light were used to determine the best-fitting peak magnitudes and fade rates (shown as dotted lines) for each eruption. The photometric data are available in the appendix, Table~A1.
}
\label{fig:f1}
\end{figure*}

\section{The Light Curves}

Figure~\ref{fig:f1} shows the light curves for 2020-01b and 2022-09a, the last two observed eruptions of M31N 1926-07c. It is clear that the light curves of the two eruptions are quite similar, with peak observed magnitudes of $R=17.2\pm0.1$ and $CR\footnote{
Since the BGO $CR$ photometry is $\sim$0.2 magnitudes brighter on average compared with our $R$ band data, we have estimated that the peak $R$-band magnitude of the 2022 eruption is given by $R=17.1\pm0.1$.}
=16.9\pm0.1$ being reached during the 2020 and 2022 eruptions, respectively. The fade rates also appear to be reasonably consistent, however
the 2022 eruption seems to have faded somewhat more quickly at late times ($\grtsim8$ days post peak).

If we restrict our analysis to the first two weeks after peak brightness when the photometry was most reliable,
linear least-squares fits to the declining branches of the 2020 and 2022 light curves give fade rates of $0.1909\pm0.0145$ and $0.1959\pm0.0094$ mag per day, which correspond to $t_2(R)=10.5\pm0.8$ and $t_2(R)=10.2\pm0.5$~days, respectively, in excellent agreement.
The predicted peak magnitudes are also very similar, where fits to the 2020 and 2022 light curves give, respectively, $R_\mathrm{peak}=17.4\pm0.06$ and $R_\mathrm{peak}=17.5\pm0.04$.

Given that the best-fitting peak magnitudes slightly underestimate the observed peak brightnesses, it may be that our measured fade rates slightly overestimate the $t_2$ times. In keeping with the precise definition of $t_2$ -- the time for a nova to fade 2 magnitudes below {\it peak\/} brightness -- we have also determined $t_2$ following this prescription, finding slightly shorter times of $t_2(R)=9.7\pm0.9$ and $t_2(R)=8.1\pm0.5$~days for the 2020 and 2022 eruptions, respectively.

\section{Discussion}

Recurrent novae with very short intervals between successive eruptions ($t_\mathrm{recur}\lessim3$~yr) are thought to harbor massive white dwarfs ($M_\mathrm{WD}\grtsim1.2~M_{\odot}$) that accrete at a high rate ($\dot M \grtsim 3\times10^{-7}~M_{\odot}~\mathrm{yr}^{-1}$) before achieving the accreted mass ($\sim10^{-6}~M_{\odot}$) necessary to trigger a thermonuclear runaway \citep[e.g., see][]{2014ApJ...793..136K}. The low ejected mass is expected to result in a rapid photometric evolution characterized by a short $t_2$ time. For example, M31N 2008-12a with its one year
recurrence time has a $t_2$ time of only $\sim$2~days \citep{2015A&A...580A..45D}.

Although M31N 1926-07c is considered a ``fast" nova, its $t_2$ time of $\sim$9 days is perhaps longer than one would predict for a recurrent nova with such a short inter-outburst interval.
Spectroscopic observations of the 2008-08b eruption \citep{2008ATel.1703....1D,2010AN....331..197D} show a He/N spectrum that is typical of recurrent novae, but the lines were reported to be unusually narrow (FWHM $\lessim850$ km~s$^{-1}$), perhaps consistent with the longer than expected $t_2$ time.

M31N 1926-07c with its five observed outbursts is one of the handful of recurrent novae in M31 with more than three detected eruptions. Only M31N 1963-09c and 2008-12a with their 6 and 14 outbursts, respectively, have been seen to erupt more frequently. Based on the times of previous outbursts, the next eruption of M31N 1926-07c is predicted for June 2025. It will be important to secure both photometric and spectroscopic observations at that time in order to further characterize the properties of this intriguing recurrent nova.


\vspace{5mm}
\facilities{Ond\v{r}ejov Observatory (OND) 0.65-m, Ningbo Bureau of Education and Xinjiang Observatory Telescope (NEXT) 0.6-m, Burke-Gaffney Observatory (BGO) 0.6-m, Mount Laguna Observatory (MLO) 1.0-m.}

\bibliography{M31N1926-07c}{}
\bibliographystyle{aasjournal}

\newpage
\appendix

\startlongtable
\begin{deluxetable}{cccr}
\tablenum{A1}
\tablecolumns{4}
\tablecaption{Nova Photometry}
\tablehead{\colhead{(JD - 2,450,000)} & \colhead{Filter\tablenotemark{a}} & \colhead{Mag} & \colhead{Observatory\tablenotemark{b}}
}
\startdata
\cutinhead{M31N 2020-01b}
8876.088 & R& $>20.2$      &NEXT \cr
8876.225 & R& $>19.2$      & OND \cr
8877.048 & R& $17.2\pm 0.1$&NEXT \cr
8877.206 & R& $17.5\pm 0.2$& OND \cr
8878.086 & R& $17.6\pm 0.1$&NEXT \cr
8878.234 & R& $17.4\pm 0.2$& OND \cr
8878.250 & R& $17.5\pm 0.2$& OND \cr
8879.066 & R& $17.8\pm 0.2$&NEXT \cr
8879.460 & R& $18.0\pm 0.3$& OND \cr
8880.108 & R& $18.1\pm 0.2$&NEXT \cr
8880.334 & R& $18.1\pm 0.2$& OND \cr
8881.074 & R& $18.2\pm 0.2$&NEXT \cr
8882.045 & R& $18.1\pm 0.2$&NEXT \cr
8883.068 & R& $18.6\pm 0.2$&NEXT \cr
8885.067 & R& $19.4\pm 0.3$&NEXT \cr
8885.247 & R& $18.8\pm 0.3$& OND \cr
8886.067 & R& $19.0\pm 0.4$&NEXT \cr
8887.061 & R& $19.0\pm 0.5$&NEXT \cr
8888.220 & R& $19.2\pm 0.3$& OND \cr
8888.244 & R& $19.1\pm 0.3$& OND \cr
8892.274 & R& $19.7\pm 0.4$& OND \cr
8895.299 & R& $>19.6$      & OND \cr
\cutinhead{M31N 2022-09a}
9830.376 & R& $>20.0$      & OND \cr
9830.426 & R& $19.4\pm 0.3$& OND \cr
9830.450 & R& $18.2\pm 0.2$& OND \cr
9830.472 & R& $18.0\pm 0.2$& OND \cr
9830.493 & R& $17.5\pm 0.1$& OND \cr
9830.725 & L& $16.9\pm 0.1$& BGO \cr
9831.277 & C& $17.6\pm 0.1$&NEXT \cr
9831.308 & R& $17.8\pm 0.2$& OND \cr
9831.593 & R& $17.7\pm 0.1$& OND \cr
9832.334 & C& $17.8\pm 0.1$&NEXT \cr
9832.749 & L& $17.7\pm 0.2$& BGO \cr
9833.273 & C& $18.1\pm 0.2$&NEXT \cr
9833.654 & L& $18.0\pm 0.2$& BGO \cr
9834.292 & R& $18.4\pm 0.2$& OND \cr
9834.308 & R& $18.3\pm 0.1$& OND \cr
9834.343 & R& $18.1\pm 0.2$& OND \cr
9834.571 & L& $18.0\pm 0.2$& BGO \cr
9835.301 & C& $18.1\pm 0.2$&NEXT \cr
9835.437 & R& $18.4\pm 0.2$& OND \cr
9835.462 & R& $18.4\pm 0.2$& OND \cr
9835.678 & L& $18.2\pm 0.3$& BGO \cr
9836.314 & R& $18.5\pm 0.1$& OND \cr
9836.335 & R& $18.5\pm 0.1$& OND \cr
9837.579 & L& $18.6\pm 0.3$& BGO \cr
9837.754 & R& $18.8\pm 0.2$& MLO \cr
9838.258 & C& $19.3\pm 0.4$&NEXT \cr
9839.758 & R& $19.6\pm 0.2$& MLO \cr
9840.337 & R& $19.4\pm 0.3$& OND \cr
9840.774 & R& $19.7\pm 0.3$& MLO \cr
9841.759 & R& $19.9\pm 0.3$& MLO \cr
9843.746 & R& $19.9\pm 0.2$& MLO \cr
9844.422 & R& $19.9\pm 0.5$& OND \cr
9845.267 & R& $20.1\pm 0.4$& OND \cr
9845.761 & R& $21.4\pm 0.5$& MLO \cr
9845.313 & R& $20.5\pm 0.4$& OND \cr
9846.741 & R& $21.9\pm 0.5$& MLO \cr
9847.739 & R& $21.9\pm 0.5$& MLO \cr
9848.734 & R& $22.2\pm 0.7$& MLO \cr
\enddata
\tablenotetext{a}{R -- Johnson $R$; L -- Luminance; C -- Clear (no filter)}
\tablenotetext{b}{OND: Ond\^{r}ejov -- 0.65-m; NEXT: Ningbo Bureau of Education and Xinjiang Observatory Telescope -- 0.6-m; BGO: Burke Gaffney Observatory -- 0.6-m; MLO: Mount Laguna Observatory -- 1.0-m}
\end{deluxetable}

\end{CJK*}
\end{document}